\documentclass[twocolumn,preprintnumbers,amsmath,amssymb, superscriptaddress]{revtex4-2}
\usepackage{graphicx}
\usepackage{xcolor}
\usepackage[english]{babel}

\begin{document}
\preprint{}

\title{Erbium-based multifuncional compounds as molecular microkelvin-tunable driving-sensing units.}


\def \FUW{Institute of Experimental Physics, Faculty of Physics, University
of Warsaw, Pasteura 5, 02-093 Warsaw, Poland}
\def \CFis{CFisUC, Physics Department, Universidade de Coimbra, Rua Larga, P-3004-516, Coimbra, Portugal.}
\def \EPS{Advanced Materials Laboratory, ETSIIAA, Universidad de Valladolid, Avenida de Madrid 44, 34004 Palencia, Spain.}
\def \MALTA{Department of Physics and MALTA Consolider Team, Universidad de La Laguna, E-38206 San Cristóbal de La Laguna, Santa Cruz de Tenerife, Spain. }

\author{Jaros{\l}aw Rybusinski} \affiliation{\FUW}
\author{Tomasz Fas} \affiliation{\FUW}
\author{Pablo Martin-Ramos} \affiliation{\EPS}
\author{Victor Lavín} \affiliation{\MALTA}
\author{Jacek Szczytko} \affiliation{\FUW}
\author{Jan Suffczynski} \affiliation{\FUW}
\author{Inocencio R. Martín} \affiliation{\MALTA}
\author{Jesus Martin-Gil} \affiliation{\EPS}
\author{Manuela Ramos Silva} \affiliation{\CFis}
\author{Bruno Cury Camargo}\email[]{b.c_camargo@yahoo.com.br} \affiliation{\FUW} 
 
\date{\today}

\begin{abstract} 
We demonstrate the selective control of the magnetic response and photoluminescence properties of $\text{Er}^{3+}$ centers with light, by associating them with a highly conjugated $\beta$-diketonate (1,3-di(2-naphtyl)-1,3-propanedione) ligand. We demonstrate this system to be an optically-pumped molecular compound emitting in infra-red, which can be employed as a precise heat-driving and detecting unit for low temperatures.
\end{abstract}

\maketitle

\section{Introduction}\label{sec:Intro}

Lanthanide ion coordination compounds possess fascinating physical properties and important technological applications, which are often governed by their magnetism and optical responses \cite{Binnemans2009, Katkova2010, Errulat2019}. Unfortunately, such ions usually present poor  absorption characteristics, causing a severe hindrance to their optically-pumped luminescence. Luckily, the photo-excitation of magnetically-relevant emitting levels in these materials can be achieved almost at will by employing organic ligands as chromophores. These components strongly absorb light at selected wavelengths, sensitizing the magnetic ions through intramolecular energy transfer – called the ``antenna effect'' \cite{Eliseeva2010, Weissman1942}. 

$\beta$-diketones containing aromatic groups are prime candidates for such a role, as they exhibit strong absorption over a wide wavelength range and are known to provide efficient energy transfer to lanthanide ions \cite{Yang2002, Sun2005}. Among possible choices, the highly conjugated $\beta$-diketonate 1,3-di(2-naphthyl)-1,3-propanedione (Hdnm) is expected to sensitize coordinated Er(III) and Eu(III) upon excitation in the visible range ($>400$ nm) \cite{Forster1959, Dogariu1999, Kang2003, Divya2011, Ramos2016}.

However, the crystallization of organic compounds remains a challenging task, with growth attempts often suffering from low rates, low yields, and polymorphic specimens with minute fractions of the desired phase \cite{Price2018}. Optical measurements in such small quantities – tenths or hundredths of micrograms – pose a challenge in terms of proper collection and amplification of the scattered light.

An alternative is to explore thermodynamic properties at equilibrium (e.g. specific heat) of illuminated bulk samples. Conventional calorimetry, however, involves the same challenges associated with optical experiments in very small specimens, and dedicated instrumentation is required (see, e.g., ref. \cite{Doettinger2001}). On the other hand, the strong magnetic response of rare-earth-based compounds makes such materials excellent candidates for magnetic measurements. Indeed, conventional commercial SQUID magnetometry is capable of detecting the magnetic response of as little as $10^{13}$ spins, which for the typical Er-based composites considered here, corresponds to $10^{12}$ molecules ($\sim$ 4 ng).

This precision allows one to observe, a priori, variations in the magnetic response or the magnetic ion when illuminating the molecular compound. In this work, we explore this possibility by studying the magnetism of $[\text{Er(dnm)}_3\text{(bipy)}]$ subjected to light excitation. We demonstrate that SQUID magnetometry in this strong paramagnet acts as a viable thermodynamical alternative to probe the optical and thermal properties of minute sample quantities. Possible applications of such a system in thermometry are discussed.

\section{Samples and experimental setups \label{sec:Methods}}

\begin{figure}[ht]
    \centering
    \includegraphics[width=1\linewidth]{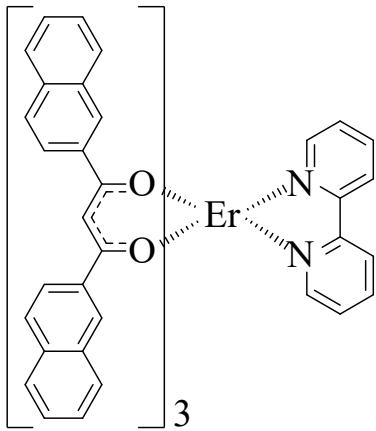}
    \caption
    {\label{fig:1} Chemical structure of $\text{Er(dnm)}_3\text{(bipy)}$.}
\end{figure}

The sample considered in this study is tris(1,3-di(2-naphthyl)-1,3-propanedionate)mono(2,2'-bipyridine)erbium(III), [$\text{Er(dnm)}_3\text{(bipy)}$, $\text{C}_{79}\text{H}_{53}\text{ErN}_2\text{O}_6$]. This novel compound, in a form of a yellow powder, was synthesized in-house following the procedure outlined in the supplementary information (SI) \cite{suppl_material}. Its structure consists of a central rare-earth element ($\text{Er}^{3+}$), surrounded by six oxygen atoms from $\beta$-diketonate ligands and two nitrogen atoms from a 2,2'-bipyridine neutral molecule (see Fig. \ref{fig:1}), prompting a distorted square antiprism chemical environment for the lanthanide.  Chemical and structural analysis confirmed a pure, yet disordered, solid (see the SI for the full characterization \cite{suppl_material}).

After the synthesis, the physical properties of the  material were probed through optical and magnetic measurements. Optical characterization was achieved through absorption and luminescence measurements in the 5.0 K $\leq$ T $\leq$ 300.0 K temperature interval. They were performed by illuminating the sample with the laser light in a wavelength range between 400 nm and 800 nm, with a linewidth of 2 nm, using a LLTF Contrast powered by NKT 8 W supercontinuum laser. The scattered and emitted light responses were captured by an Andor SR-500i spectrometer with an IDus InGaAs array. Magnetic measurements were carried out on a QuantumDesign MPMS 7T platform, in the  2 K $\leq$ T $\leq$ 300 K temperature range, under magnetic fields up to 7 T \cite{QD_INC}. The setup was adapted with an optic fiber window, to allow illumination of the sample during magnetic measurements. For this purpose, a spectrometer-filtered Xenon arc discharge light bulb was employed.

\section{Results \label{sec:Results}}

\begin{figure}[ht]
    \centering
    \includegraphics[width=1\linewidth]{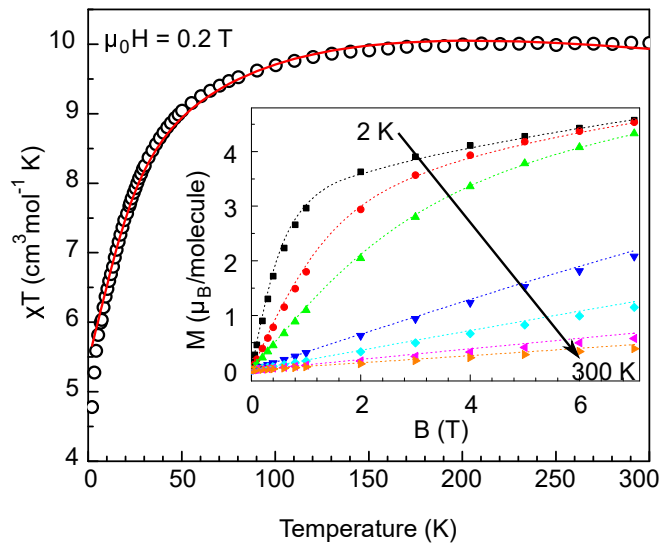}
    \caption
    {\label{fig:2} $\chi$T vs. temperature for the sample considered herein, obtained at $\mu_0H = 0.2$ T. The inset shows the magnetic response per molecule measured at (from top to bottom) T = 2 K, 5 K, 10 K, 50 K, 100 K, 200 K and 300 K. The red line in the main panel and the dashed lines in the inset represent magnetization curves obtained with Phi \cite{Chilton2013} for an isolated $J = 15/2$ $\text{Er}^{3+}$ magnetic ion in a distorted $D_{4d}$ crystallographic environment. The y-axis experimental uncertainty is not visible in the scale shown.}
\end{figure}

Temperature-scaled magnetic susceptibility vs. temperature ($\chi\text{T}\times \text{T}$) and magnetization vs. magnetic field ($\text{M}\times\text{H}$) for the compound under study are shown in Fig. \ref{fig:2}. Measurements revealed a clear paramagnetic-like behavior, which was, however, not well-described by the conventional Curie-law ($\chi \propto \text{T}^{-1}$)  above 6 K. This can be attributed to the chemical environment for the $\text{Er}^{3+}$ ion in $[\text{Er(dnm)}_3\text{(bipy)}]$. Indeed, simulations using PHI software \cite{Chilton2013} reproduced well the $\text{M}\times\text{H}$ and $\chi\text{T}\times \text{T}$ sample behavior by assuming isolated $J = 15/2$ magnetic centers in a $D_4$ crystalline environment, in agreement with XRD data (see the SI \cite{suppl_material}).

\begin{figure}[ht]
    \centering
    \includegraphics[width=1\linewidth]{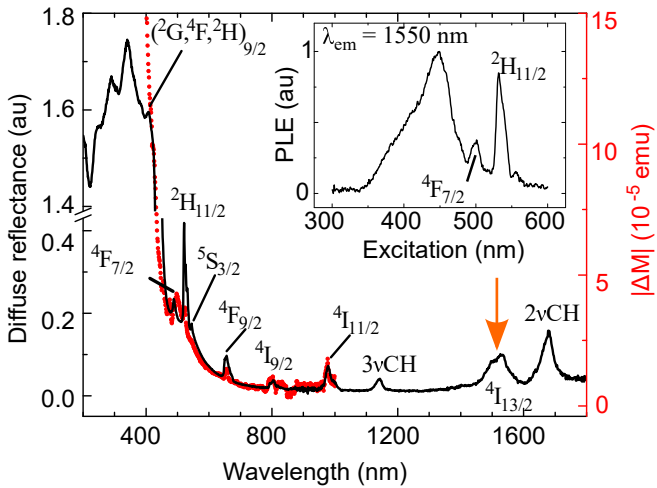}
    \caption
    {\label{fig:3} Diffuse reflectance (black line, left axis) and variation of magnetic response measured at T =  2 K and $\mu_0H = 0.2$ T  ($|\Delta\text{M}| \equiv |\text{M}_\text{dark}-\text{M}_\text{lit}|$, red points, right axis) as a function of wavelength. The labeled absorption lines correspond to transitions from the ${}^{4}{\text{I}}_{15/2}$ ground state of the $\text{Er}^{3+}$ ion, while the smooth background is associated with the organic ligand. The inset shows the photoluminescence measured at $\lambda = 1550$ nm (pointed by an arrow in the main panel). The labelled peaks in the inset are associated with the corresponding Er absorption lines of the main panel, while the broad maximum centered at $\lambda \approx 450$ nm is due to the energy transfer from the organic ligand to the metallic ion.}
\end{figure}

Strikingly, upon illumination, a remarkable variation of the sample magnetic response was observed. Its magnitude, denoted $|\Delta\text{M}| \equiv |\text{M}_\text{dark}-\text{M}_\text{lit}|$,  was determined by performing consecutive magnetization measurements for the sample without ($\text{M}_\text{dark}$) and under ($\text{M}_\text{lit}$) irradiation at fixed temperatures. The amplitude of $\Delta\text{M}$ closely followed the intensity profile of the diffuse reflectance spectrum of the material, and was most pronounced at low temperatures (see Fig. \ref{fig:3}). 

Among the features visible both in magnetic and optical measurements, a broad absorption band in the 200-500 nm range can be mainly attributed to the $\pi$-$\pi^*$ transitions of the dnm $\beta$-diketonate \cite{Sun2013}, with overlapping bands from the 2,2'-bipyridine organic ligand (240-290 nm region) \cite{Ramos2015} and from the ${}^{4}{\text{I}}_{15/2} \rightarrow ({}^{2}\text{G},{}^{4}\text{F},{}^{2}\text{H})^{9/2}$ $\text{Er}^{3+}$ transition (407 nm). Above 490 nm, sharp peaks are associated with intra-configurational ${}^{4}f^{11}-{}^{4}f^{11}$ electronic transitions starting from the ${}^{4}{\text{I}}_{15/2}$ ground state of the $\text{Er}^{3+}$ magnetic center, superimposed to the ligand's absorption tail \cite{Meer1994}.

The excitation spectrum of the transition at $\lambda \approx 1550$ nm (${}^{4}{\text{I}}_{13/2} \rightarrow {}^{4}{\text{I}}_{15/2}$, see the inset in Fig. \ref{fig:3}) featured a broad peak when the sample was pumped with a wavelength of about 450 nm. Such a feature is significantly red shifted compared to other rare-earth-based coordination compounds \cite{Kang2003, Sun2013}, including those based on $\beta$-diketonate complexes \cite{Ramos2015}. Its occurrence is closely related to the absorption maximum of the organic ligand used here, which occurs in the UV-Vis range (see Fig. \ref{fig:3} and the SI \cite{suppl_material}). Such a result strongly indicates the sensitization of the emissive metal center in our compound by the antenna effect.

\begin{figure}[ht]
    \centering
    \includegraphics[width=1\linewidth]{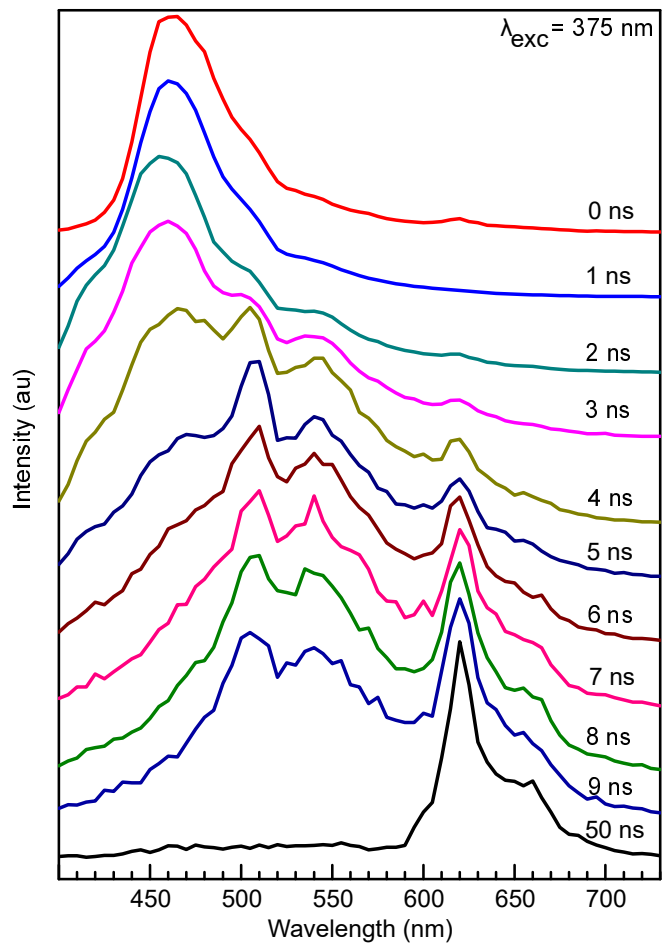}
    \caption
    {\label{fig:4} Time-resolved evolution of the PL emission in the visible range of $\text{Er(dnm)}_3\text{(bipy)}$.}
\end{figure}

To better understand how absorption through the ligand influenced on the magnetic center of the molecule, we measured the temporal evolution of the sample's photoluminescence at excitation pulses with wavelength $\lambda_{\text{exc}} = 375$ nm. This value is centered around the broad UV absorption band of the $\text{(dnm)}_3$ ligand (see Fig. \ref{fig:3}). The results revealed that an initially strong fluorescent emission at $\lambda \approx 480$ nm quickly gave way to an emission line centered at $\lambda \approx 620$ nm (see Fig. \ref{fig:4}). The latter is ascribed to the triplet state of the organic ligand, and largely overlaps with the ${}^{4}\text{F}_{9/2}$ absorption line of $\text{Er}^{3+}$ (for a full description, see the SI \cite{suppl_material}). This results in the pumping of the rare earth ion through a resonant energy transfer process \cite{Meer1994}, followed by relaxation and radiative decay (see the diagram provided in the SI \cite{suppl_material}, Fig. S7), thus yielding the characteristic luminescence spectra at around 1550 nm showcased in the inset of Fig. \ref{fig:3}.

The triplet state at $\lambda \approx 620$ nm exhibited a non-exponential decay, with characteristic lifetimes of a few nanoseconds (see the SI \cite{suppl_material}, Sec. V). These relatively small values suggest an efficient ligand-to-metal energy transfer in the system \cite{Jia2020}. The PL decay of the $\text{Er}^{3+}$ ${}^{4}\text{I}_{13/2}$ multiplet ($\lambda \approx 1550$ nm), however, exhibited a single exponential behavior, which indicates a consistent coordination environment around the lanthanide ion (see the SI \cite{suppl_material}, Fig. S9). The characteristic lifetime extracted for this transition ranged around $\tau \approx 1.3$ $\mu \text{s}$, comparable to other $\text{Er}^{3+}$ compounds found in the literature \cite{Ramos2015, Li2008, Li2009}. Nevertheless, considering the radiative lifetime of $\text{Er}^{3+}$ at approx. 1-2 ms allows the estimation of the quantum efficiency of the transition at $\sim 0.1$\% \cite{Polman2001}. This suggests that most energy captured by the organic ligands is absorbed by the material, rather than being re-emitted by the rare earth center.

Such a small quantum efficiency does not allow $\Delta\text{M}(\lambda)$, shown in Fig. \ref{fig:3}, to be mainly attributed to a variation of the magnetic state of the $\text{Er}^{3+}$ ion. Indeed, assuming that each incident photon is absorbed by a molecule, leasing to a change of the magnetic state in the Er ion, the required power delivery at $\lambda = 520$ nm to induce $\Delta \text{M} \approx 10^{-5}$ emu at T = 2 K would be of $\approx 4$ W. This a value is unrealistic in our setup. Instead, at the highest applied power, the population of excited Er ions in dynamical equilibrium is not larger than $10^8$ (assuming a long, 10~$\mu$s relaxation time). This is far below the detection threshold of $10^{18}$ spins for the SQUID magnetometer. 

\begin{figure}[ht]
    \centering
    \includegraphics[width=1\linewidth]{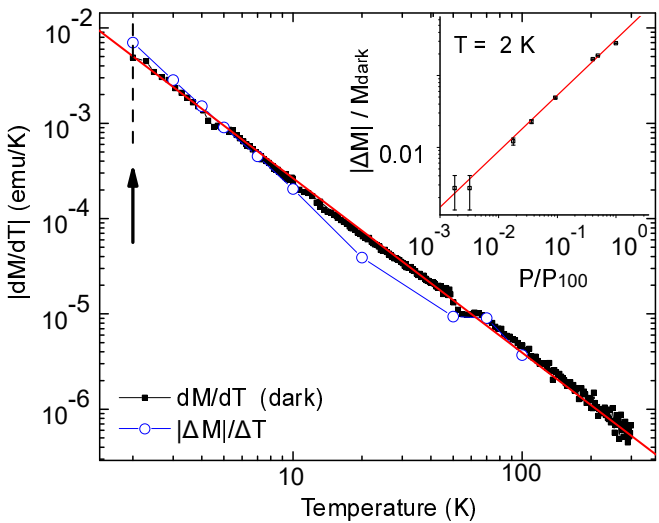}
    \caption
    {\label{fig:5} $d\text{M}/d\text{T}$ vs. temperature (filled black squares), plotted together with  $|\Delta\text{M}|/\Delta \text{T}$ (empty blue circles). In it, $|\Delta\text{M}|$ is defined as  $|\Delta\text{M}| \equiv |\text{M}_\text{dark}-\text{M}_\text{lit}|$, where $\text{M}_\text{lit}$ was measured for a sample illuminated with $P_{100}=1170$ $\mu\text{W}$ of light at $\lambda = 520$ nm. The term $\Delta\text{T}$ is a scaling parameter set at $\Delta\text{T} = 0.25$ K, chosen so that both curves would overlap. The red line is a function of the type $\Delta \text{M}' = \text{M}'_0 (\text{T}/\text{T}_0)^{-1.8}$, $\text{T}_0 = 1$~K and $\text{M}'_0 = 18\times 10^{-3}$ emu/K. Error bars are not visible in the scale. The inset shows the variation rate $\Delta \text{M}/\text{M}_{dark}$ measured at $\text{T} = 2$~K as a function of the relative light intensity $P/P_{100}$ (region indicated by a black arrow and a dashed line in the main panel). The red line in the inset follows the relation $| \Delta \text{M} | / \text{M}_{dark} \propto (P/P_{100})^{0.8}$.}
\end{figure}

Instead, the change in the magnetic response upon illumination was directly correlated with the change of M with T. This is shown in Fig. \ref{fig:5} for the brightest excitation available in our experimental setup, which was of $P_{100} = 1170$ $\mu\text{W}$ at the sample location. In the figure, a remarkable overlap is observed between $d\text{M}/d\text{T}$ and $|\Delta \text{M}|/\Delta \text{T}$, with $\Delta \text{T}$ a constant representing a temperature change. This result strongly suggests heat as the driving mechanism behind the modulation of the sample magnetization, following $\Delta \text{M} = d\text{M}/d\text{T} \times \Delta \text{T}$. For the case shown in the figure, the variation in sample temperature is estimated at $\Delta \text{T} \approx 0.25$ K. Such a value could be indeed controlled by changing the intensity of the incident light. This is illustrated in the inset of Fig. \ref{fig:5}, through the demonstration that $|\Delta\text{M}|$ was well-described by the phenomenological relation $\Delta \text{M} \propto (P/P_{100})^{0.8}$ over the  $2.1 \text{ }\mu\text{W }\leq P \leq 1170$~$\mu\text{W}$ interval. 

Nonetheless, it should be noted that the power delivered to the sample can be continuously tuned by controlling the excitation wavelength, and not only its strength. This occurs because the absorption background of the molecule is non-monotonic (see Fig. \ref{fig:3}). That is, the compound is heated more intensely for UV wavelengths and around absorption lines. For $\lambda \approx 520$ nm, the relation reads $\Delta \text{T} = 0.25 \times (P/P_{100} )^{0.8}$~K. It should be stressed that this possibility results from the choice of a ligand with a broad-band energy-absorption, bound to a rare-earth magnetic ion. The former acts as the heating element of the molecule, whereas the latter provides a mensurable temperature-dependent quantity.

Knowing the $\text{M}_\text{dark}(\text{T})$ behavior of the sample thus allows a fine control of the system's temperature through tracking the magnetic response of the material in the presence of light. Considering a realistic resolution of the SQUID magnetometer utilized here at $10^{-7}$~emu, allows a precision of the temperature driving system at $\Delta \text{T}_{\text{min}} \approx 10^{-7}\text{ emu} \times(d\text{M}/d\text{T})^{-1}$ . For the compound chosen, $d\text{M}/d\text{T} \propto \text{T}^{-1.8}$ (see Fig. \ref{fig:5}), which translates to an enhanced resolution at lower temperatures. In the present work, the value attained in $\Delta \text{M}(\lambda,\text{T}=  2\text{ K})$ allows to infer a resolution $\Delta \text{T}_{\text{min}} \approx 20$ $\mu\text{K}$. The mass of the sample (0.2 mg) sets the sensitivity of our sample/heating device at 100 $\mu\text{K}$/mg. However, this sensitivity is likely to be at least two orders of magnitude better, as absorption occurs in the sample surface region over a depth of $\sim$100 nm, which may be estimated at not more than 10\% of the total volumetric content.

Considering the scenario outlined above, it is feasible to embed this magnetic compound in a matrix of interest, and track the temperature of the latter by measuring the magnetic properties of the Er complex. Provided that the molecular compound remains in a dilute regime,  is chemically passive in relation to the matrix, and that crystalline fields introduced by the matrix (if any) remain weak in comparison with local bonds, no major variations on the magnetic response are expected. This occurs because the main factor dictating magnetic properties of weakly interacting molecular compounds is the ligand field around the magnetic center, which is sensitive to the molecular configuration \cite{Book_Kahn}.

Similar approaches to temperature sensing have been employed in the past, e.g. through the use of the magnetic response of magnetite \cite{Arrott1981} and paramagnetic salts \cite{Fu1998}, or through tracking the photoluminescence response of organic-rare-earth compounds \cite{Errulat2019}. Other applications have also been suggested for Er-based materials (see e.g. \cite{Polman2001}), relying on their spectroscopic properties. However, here we combine both the optical and magnetic properties, resulting in a sensing-driving device for fine temperature control. We stress that the choice of a hybrid Er-organic compound poses a decisive advantage over the simple inclusion of magnetic centers as thermometers (as in the case of magnetite \cite{Arrott1981}, paramagnetic salts \cite{Fu1998}, or single Er ions \cite{Polman2001}). In addition to the possibility of passivizing the magnetic ion with respect to the matrix of interest, the organic ``antenna'' of these molecules can also have absorption lines engineered at wavelengths for which the matrix is transparent \cite{Sun2005, Yang2002}. This ensures the delivery of thermal power to the system at the detecting centers (rather than at the sample surface), which can be (for example) diluted across the material to ensure homogeneous heating/sensing. We also note that the resolution reported by us depends on the magnetic response of the material under consideration, and may be altered by engineering a desired MxT slope.

Taking into consideration the relative flexible operational temperature range, the possibility of a tunable optical excitation, and the relatively simple synthesis method for the material shown here, our reported resolution of 100~$\mu$K/mg is competitive in relation with other existing alternatives with detection limits in the $\mu$K range. Among them, state-of-art low-temperature thermometry using on-chip Coulomb blockade sensors yields a precision of 1\% for temperatures around 10 mK \cite{Sarsby2020}. Albeit very powerful, such a technique comes at a much larger cost, requires electrodes, is confined to low temperatures, and demands complex microfabrication processes. Another interesting approach makes use of an infra-red pyrometer operating in a differential configuration on thermally-stable surroundings. Such an instrument is reported to reach a remarkable precision around 30~$\mu$K at room temperature, but is a passive technique fit for chemical and biological processes \cite{Bai2020}. Finally, opto-mechanical measurements probing vibrating modes of nanometric membranes yield resolutions around 15~$\mu$K \cite{Ferreiro2021}. Such impressive values, however, are suitable for on-chip detection, as they are inherently sensitive to the environment surrounding the membranes. The method outlined here by us, in contrast, requires a small magnetic field to probe temperature, but is contact-free and is based on a compound that may be dispersed in the material being probed. This approach eliminates electronic heat contribution that may disrupt measurements otherwise (see e.g. \cite{Xing2021}) and ensures thermalization with the sample. Such an approach is promising for systems in which electrodes are unnecessary, as it minimizes conductive heat transfer while allowing selective heating by tuning light excitation wavelength at a fixed output power.

\section{Conclusions \label{sec:Conclusions}}

In summary, we have established the magnetic and optical properties of a newly-synthesized magnetic molecule based on a $\beta$-diketonate ligand attached to a rare-earth $\text{Er}^{3+}$ ion. The presence of the organic ligand in this compound acts as an absorption center (or ``antenna'') for the incoming radiation. Part of the absorbed energy is transferred towards the $\text{Er}^{3+}$, which causes the material to act as an optically-pumped IR-emitting compound. The remaining energy is absorbed by the material, which then acts as a wavelength-tunable heating source. Its temperature can be minutely tracked through the strong paramagnetic response of the embedded magnetic ion.  The resolution of the temperature tracking using this method reciprocally increases with the temperature, reaching the value of 20 $\mu\text{K}$ at $\text{T} = 2$ K for the compound shown here. By properly choosing the organic antenna attached to the $\text{Er}^{3+}$ ion, the engineering of local heater/thermometers embedded in a sample of interest may be envisaged. The system can be homogeneously driven at wavelengths for which the sample is transparent, while simultaneously having its temperature magnetically probed.

\section*{Acknowledgements}
This work was carried out with the support of FCT (Fundação para a Ciência e a Tecnologia)  UIDB/04564/2020, and UIDP/04564/2020.

\section*{Data availability}
The data that support the findings of this study are available from the corresponding author upon reasonable request.

%

\end{document}